\documentclass[%
 reprint,
 amsmath,amssymb,
 aps,
]{revtex4-1}

\usepackage{graphicx}
\usepackage{dcolumn}
\usepackage{bm}

\begin{document}

\preprint{APS/123-QED}

\title{Stellar structure model in hydrostatic equilibrium in the context of $f(\mathcal{R})$-gravity}

\author{Ra\'ila Andr\'e}
 \altaffiliation{andreraila@gmail.com}
\author{Gilberto M. Kremer}%
 \email{kremer@fisica.ufpr.br}
\affiliation{%
 Departamento de F\'isica, Universidade Federal do Paran\'a, 81531-980 Curitiba, Brazil.\\
}%


\date{\today}

\begin{abstract}
In this work we present a stellar structure model from $f(\mathcal{R})$-gravity point of view capable to describe some classes of stars (White Dwarfs, Brown Dwarfs, Neutron stars, Red Giants and the Sun). This model was based on $f(\mathcal{R})$-gravity field equations for $f(\mathcal{R})= \mathcal{R}+ f_2 \mathcal{R}^2$, hydrostatic equilibrium equation and a polytropic equation of state. We compared the results obtained with those found by the Newtonian theory. It has been observed that in these systems, where high curvature regimes emerge, stellar structure equations undergo modifications. Despite the simplicity of this model, the results were satisfactory. The estimated values of pressure, density and temperature of the stars are within those determined by observations. This $f(\mathcal{R})$-gravity model has been proved to be necessary to describe stars with strong fields such as White Dwarfs, Neutron stars and Brown Dwarfs, while stars with weaker fields, such as Red Giants and the Sun, were best described by the Newtonian theory. 
\end{abstract}

\pacs{Valid PACS appear here}
\maketitle


\section{\label{sec:level1}Introduction}

The $f(\mathcal{R})$-gravity is a class of theories that represent an approach to gravitational interaction. In this context, General Relativity (GR) has to be extended in order to solve several issues. This approach considers modifications of the Einstein-Hilbert action in order to include higher-order curvature invariants with respect to the Ricci scalar \cite{schmidt2007, sotiriou2010}. From the astrophysical and cosmological point of view, the objective is to explain phenomena such as dark energy and dark matter under a geometric pattern \cite{capozziello2004, capozziello2005proceeding, capozziello2007, martins2007, boehmer2008, nojiri2007a, battye2015, hu2007, tsujikawa2010} with the possibility that gravitational interaction depends on the scales. In this sense, in principle, these theories do not require the introduction of new particles and preserve all sucessful results of Einstein's theory, based on the same fundamental physical principles. It is also known that some $f(\mathcal{R})$ gravity models can pass in test performed in the weak-field solar system \cite{hu2007}. In addition, a considerable number of viable $f(\mathcal{R})$ models are known, among which we highlight \cite{nojiri2003,nojiri2007b, faulkner2007, faraoni2006, cognola2008}.
Futhermore, no extended gravity model, until this moment, can handle all phenomenology ranging from the quantum scale to the cosmological scale \cite{nojiri2011}. Another problem is that the description of the $f(\mathcal{R})$ theory is substantially equivalent to the description associated with the hypothesis of dark components. This interpretation arises from the fact that the degrees of freedom present in the $f(\mathcal{R})$ theory can be expressed by an effective energy-momentum tensor able of giving rise to the dark matter effects \cite{capozziello2012, sotiriou2010}. From this picture emerges the necessity of observations capable of preserving or excluding one of these theories.

In this framework, several works related to relativistic stars considering extended gravity theories have been developed. In \cite{cooney2010}, using the method of perturbative constraints and corrections of the form $\mathcal{R}^{n+1}$, they showed that the predicted mass-radius relation for neutron stars differs from that calculated in the General Relativity. In \cite{capozziello2016}, the mass-radius diagram for static neutron star were obtained in $f(\mathcal{R})$ gravity for two functions: $f(\mathcal{R})= \mathcal{R} + \alpha \mathcal{R}^2 (1 + \gamma \mathcal{R})$ and $f(\mathcal{R})= \mathcal{R}^{1+\epsilon}$. New terms related to curvature corrections emerged and modified the evolution of the mass-radius relation. In \cite{astashenok2017}, realistic models of relativistic stars in $f(\mathcal{R}) = \mathcal{R} + \alpha R^2$ gravity have been explored. In this context, the authors presented a study on the existence of neutron and quark stars for various $\alpha$ with no intermediate approximation in the system of equations. On the other hand, in \cite{astashenok2015a}, quark star models with realistic equation of state in nonperturbative $f(\mathcal{R})$ gravity have been considered. The authors showed that it is possible discriminate modified theories of gravity from General Relativity due to the gravitational redshift of the termal spectrum emerging from the surface of the star. In \cite{arapoglu2016}, a stellar structure model in $f(\mathcal{R}) = \mathcal{R} + \alpha \mathcal{R}^2$ theory was considered, using the method of matched asymptotic expansion to handle the higher order derivatives in field equations. Solutions were found for uniform density stars matching to the Schwarzschild solution outside the star. The mass-radius relations were obtained, observing the dependence of maximium mass on $\alpha$. In \cite{astashenok2013}, neutron star models in perturbative $f(\mathcal{R})$ have been considered with realistic equations of state. The mass-radius relations for $f(\mathcal{R}) = \mathcal{R} + \beta \mathcal{R} \Big[exp (\frac{- \mathcal{R}}{\mathcal{R}_0}) - 1\Big]$ and $\mathcal{R}^2$ models with logarithmic and cubic corrections were obtained. In the case of cubic corrections, stable star configurations at high central density were obtained. Such an effect could give rise to more compact stars than in General Relativity. In \cite{alavirad2013}, considering a logarithmic $f(\mathcal{R})$ theory, the authors showed that the model exhibts a chameleon effect which completely eliminates the effect of the modification on scale exceeding a few radii, but close to the surface of the neutron star, the deviation from General Relativity can significantly affect the surface redshift. In \cite{astashenok2014}, the authors showed that for a simple hyperon equation of state it is possible to obtain the maximal neutron star mass (which satisfies the recent observational data) in higher-derivative models with power-law terms as $f(\mathcal{R}) = \mathcal{R} + \gamma \mathcal{R}^2 + \beta \mathcal{R}^3$. And in \cite{astashenok2015b}, the authors studied neutron stars with strong magnetic fields. They took into account models derived from $f(\mathcal{R})$ and $f(G)$ theories where function of the Ricci curvature invariant $\mathcal{R}$ and the Gauss-Bonnet invariant $G$ are respectively considered. In this model, the maximal mass of neutron star had a considerable increasing in $f(\mathcal{R})$ with cubic corrections.

In the model discussed in this work we propose second-order corrections in the Ricci scalar adopting $f(\mathcal{R}) = \mathcal{R} +f_2 \mathcal{R}^2$. Such orders in the correction are an extension of General Relativity and are particularly interesting in cosmology, since they allow the construction of a self-consistent inflationary model \cite{starobinsky1980}. We present the results obtained for a stellar model in hydrostatic equilibrium according to the $f(\mathcal{R})$-gravity theory. The goal is to control precisely how the results deviate from those obtained through General Relativity in order to see how strong gravity regimes affect the pressure, temperature and density. We also compare these results with those related to the Newtonian theory, already known in the literature \cite{chandrasekhar1957}. 

In our model we consider a polytropic equation of state, since this equation plays an importante role in stellar structure models. This equation correctly represents the stellar gas behavior and, consequently, solves the fundamental problem of these structures together with the hydrostatic equilibrium equation. The motivation to employ the polytropic equation in the study of stellar structure is the simple nature of the polytropic structure and its correspondence with known classes of stars. Such simplicity provides a basis for the incorporation of additional effects (such as rotation), and thus an insight into the nature of the effects on true stars \cite{horedt2004}.

The interest of this model, in the present context, is manifested in the fact that, due to the expressive gravitational field, the interior of the stars can be seen as appropriate places to test alternative theories of gravity. In these regions, high curvature regimes can emerge and modify the stellar structure. In this way, we aim to show that the pressure, temperature and density can be consistently reached by the extended theories of gravity, such as $f(\mathcal{R})$-gravity for $f(\mathcal{R})= \mathcal{R} +f_2 \mathcal{R}^2$ and how the expected changes occur in the values of these quantities. 

This paper is organized as follows: In section II we derive the modified Poisson and Lane-Emden equations through the Newtonian limit of $f(\mathcal{R})$-gravity. In section III we obtain the modified stellar structure equations. In section IV we present the numerical solutions for pressure, temperature and density obtained for Neutron stars, Brown Dwarfs, White Dwarfs, Red Giant and Sun, and we compared with those results obtained by Newtonian theory. Finally, we draw our conclusions in section V. In this work we will indicate $R$ as the radius of the star and $\mathcal{R}$ as the Ricci scalar. 

\section{\label{sec:level2}The Lane-Emden equation for a model described by the $f(\mathcal{R})$ theory}

Considering a spherically symmetric self-gravitating system in equilibrium we adopt the hydrostatic equilibrium equation represented below
\begin{eqnarray} \label{eq:1}
\frac{d \phi}{d r} = \frac{1}{\rho} \frac{d p}{d r},
\end{eqnarray}
where $\rho(r)$ is the matter density, $p$ is the pressure and $\phi$ is the gravitational potential. The equation \eqref{eq:1} is a Newtonian limit of the equation resulting from the conservation of the stress-energy tensor for a perfect fluid in hydrostatic equilibrium. It can be also achieved through the Newtonian limite of the Tolman-Oppeheimmer-Volkoff equation (equation employed to describe a spherically symmetric astrophysical system in equilibrium in GR \cite{rezzolla2013, landau1987}).

We assume a polytropic equation of state
\begin{eqnarray} \label{eq:2}
p = k \rho^{\gamma},
\end{eqnarray}
where $k$ is the polytropic constant and $\gamma$ is the polytropic exponent. Then we insert equation \eqref{eq:2} into equation \eqref{eq:1}, obtaining
\begin{eqnarray} \label{eq:3}
\frac{d \phi}{ d r} = \gamma k \rho^{\gamma - 2} \frac{d \rho}{d r}.
\end{eqnarray}
For $\gamma \neq 1$, the integration of the above equation results in 
\begin{eqnarray} \label{eq:4}
\rho = \Big(\frac{\gamma - 1}{\gamma k} \Big)^{\frac{1}{\gamma - 1}} \phi^{\frac{1}{\gamma - 1}} =  \Bigg[\frac{\phi}{(n+1)k}\Bigg]^n,
\end{eqnarray}
where the chosen integration constant was $\phi = 0$ on the surface ($\rho = 0$). The constant $n$ is known as the polytropic index and is defined as $n = \frac{1}{\gamma -1}$. Through the equations \eqref{eq:2} and \eqref{eq:4}, we obtain the following expression for the pressure
\begin{eqnarray} \label{eq:5}
p = \frac{\rho \phi}{(n+1)}.
\end{eqnarray}

To describe a stellar structure model by $f(\mathcal{R})$-gravity \cite{capozziello2011book, capozziello2011report} we adopt the action represented below
\begin{eqnarray} \label{eq:6}
S = \int d^4 x \sqrt{-g} \Big[ f(\mathcal{R}) + \chi \mathcal{L}_m \Big],
\end{eqnarray}
where the Ricci scalar is only a function of metric tensor $\mathcal{R} \equiv \mathcal{R} (g)$, $f(\mathcal{R}) = \mathcal{R} + f_2 \mathcal{R}^2$ and $\chi = 8 \pi G$, with $G$ denoting the gravitational constant. It is worth mentioning that we adopt the metric $(-, + , +, +)$ and the natural units. 

By varying the action according to the metric formalism, we obtain modified Einstein field equation in $f(\mathcal{R})$ theory
\begin{eqnarray} \label{eq:7}
f'(\mathcal{R}) \mathcal{R}_{\mu \nu} - \frac{1}{2} g_{\mu \nu} f(\mathcal{R}) + (g_{\mu \nu} \square - \nabla_{\mu} \nabla_{\nu}) f'(\mathcal{R}) &=& \\ \nonumber &-& \chi T_{\mu \nu},
\end{eqnarray}
where $f'(\mathcal{R}) = \frac{d f(\mathcal{R})}{d \mathcal{R}}$ and $T_{\mu \nu}$ is the energy-momentum tensor. \\
The trace of the above field equation reads
\begin{eqnarray} \label{eq:8}
f'(\mathcal{R}) \mathcal{R} - 2 f(\mathcal{R}) + 3 \square f'(\mathcal{R}) = -\chi T^{\sigma}_{\sigma}.
\end{eqnarray}
In this model we will address the situation where the particles of the system move at a very low speed (compared to the speed of light) and the gravitational field which they are subjected is considered weak and static. Such requirements refer to the Newtonian limit \cite{weinberg1972}. In this way, writing the equations \eqref{eq:7} and \eqref{eq:8} in this limit, we have
\begin{eqnarray} \label{eq:9}
 \mathcal{R}^{(2)}_{00} + \frac{\mathcal{R}^{(2)}}{2} + \frac{1}{3 m^2} \nabla^2 \mathcal{R}^{(2)} = -\chi \rho, 
\end{eqnarray}
\begin{eqnarray} \label{eq:10}
 \Big(1 + \frac{1}{m^2} \nabla^2 \Big) \mathcal{R}^{(2)} = - \chi \rho,
\end{eqnarray}
where $m^2 = - \frac{1}{6 f_{2}}$. Considering $(1 + \frac{1}{m^2} \nabla^2 )$ as an operator, we can pass this term to the right-hand side of the equation and expand it in Taylor's series to the order term $m^{-2}$. Hence we rewrite equation \eqref{eq:10} as
\begin{eqnarray} \label{eq:11}
\mathcal{R}^{(2)} \approx - \chi \Big(1 - \frac{1}{m^2} \nabla^2 \Big) \rho.
\end{eqnarray}
It is important to emphasize here that the validity of this equation is assured for all the cases studied in this paper, since the term $\frac{1}{m^2} \nabla^2$ should always be small. This fact can be understood by observing that $m$ is directly related to the coefficient $f_2$ and that it must be small since it is a correction of GR. There are some works where bonds for this coefficient are estimated and in all cases small values are presented for this correction coefficient \cite{berry2011, aviles2013}.

Substituting the Newtonian limit of temporal component of Ricci tensor given by $\mathcal{R}^{(2)}_{00} = \nabla^2 \phi$ \cite{weinberg1972} and inserting the equations \eqref{eq:4} and \eqref{eq:11} in \eqref{eq:9}, we obtain in spherical coordinates the following equation
\begin{eqnarray} \label{eq:12}
\frac{\partial^2 \phi}{\partial r^2} &+& \frac{2}{r} \frac{\partial \phi}{\partial r} +\frac{4 \pi G}{3m^2[(n+1)k]^n} \frac{\partial^2 \phi^n}{\partial r^2} \\ \nonumber &+& \frac{8 \pi G}{3 m^2 [(n+1)k]^n} \frac{1}{r} \frac{\partial \phi^n}{\partial r} = - \frac{ 4 \pi G}{[(n+1)k]^n} \phi^n.
\end{eqnarray}
By defining dimensionless variables
\begin{eqnarray} \label{eq:13}
z = \frac{r}{\xi}, ~~\omega (z) = \frac{\phi}{\phi_c} = \Big(\frac{\rho}{\rho_c} \Big)^{\frac{1}{n}},
\end{eqnarray}
where the index $c$ refers to the center of the star and 
\begin{eqnarray} \label{eq:14}
\xi = \sqrt{\frac{[(n+1)k]^n}{4 \pi G}\phi^{(1-n)}_c},
\end{eqnarray} 
we obtain the Lane-Emden equation for this $f(\mathcal{R})$-gravity model:
\begin{eqnarray} \label{eq:15}
\frac{d^2 \omega}{d z^2} + \frac{2}{z} \frac{d \omega}{d z} + \omega^n &+& \frac{1}{3 m^2 \xi^2} \frac{d^2 \omega^n}{d z^2} \\ \nonumber &+& \frac{2}{3 m^2 \xi^2} \frac{1}{z} \frac{d \omega^n}{d z} = 0.
\end{eqnarray}
Making $m \rightarrow \infty$, i.e. $f_2 \rightarrow 0$, we recover the Lane-Emden equation for the model described by the Newtonian gravity (see \cite{capozziello2011}). Through this equation it is possible to determine the physical quantities of the system as pressure, density and temperature. Therefore, such models allows a simple description of stars and planets. 

It is worth mentioning that the modified Lane-Emden equation was previously obtained in \cite{capozziello2011} and posteriorly in \cite{farinelli2014}. In the analysis presented in these cited articles and also in this work, the field equations for $f(\mathcal{R})$ gravity (in metric formalism), the polytropic equation and the hydrostatic condition, in the Newtonian limit, were considered as a system of equations. The difference lies in the fact that in \cite{capozziello2011} and \cite{farinelli2014}, the modified Lane-Emden equation results in an integro-differential equation while in this work, it results only in a second-order differential equation. This happens due to the approach applied in the equation \eqref{eq:10}. Due this approximation, we were able to write the modified Lane-Emden equation as showen in \eqref{eq:15}. But we also verified that in both cases presents approximately the same solutions. 
\section{\label{sec:level3}The stellar structure equations according to $f(\mathcal{R})$-gravity}

Through the solution of the Lane-Emden equation we can write expressions for the physical quantities of the stellar structure such as radius, mass, temperature, matter density and pressure. Certain values of $n$ provide a description for a class of stars, for example, for $n=1$ the solution represents a Neutron star, for $n=1.5$ we have the closest solution to completely convective stars such as Red Giants and Brown Dwarfs and for $n=3$ the solution corresponds to a fully radiative star such as Sun and stars with degenerate nuclei like White Dwarfs \cite{chandrasekhar1957, hansen2004}. Searching for a star description in this work, we analyze the solutions only for the following values of $n$: $1$, $1.5$ and $3$. 

Through equations \eqref{eq:13} and \eqref{eq:14}, we obtain the radius $R= \xi z_{(n)}$ of the star given by 
\begin{eqnarray} \label{eq:16}
R = \sqrt{\frac{[(n+1)k]^n}{4 \pi G } \phi_c^{(1-n)}} z_{(n)},
\end{eqnarray} 
where $k$ is the polytropic constant and $z_{(n)}$ is the first zero of the solution. Since the boundary of the star is indicated by $\omega =0$, i.e where $z= z_{(n)}$, in the Newtonian theory we have radius of the star $R$ and the mass $M$ of the star defined through the gravitational potential as \cite{eddington1926}:
\begin{eqnarray} \label{eq:17}
R = (r)_{\omega=0}~,~~ GM = \Big(- r^2 \frac{d \phi}{d r} \Big)_{\omega =0}.
\end{eqnarray}
Through the dimensionless variables, we write
\begin{eqnarray} \label{eq:18}
R'= (z)_{\omega= 0} = z_{(n)}~,~~M'= \Big(-z^2 \frac{d \omega}{d z} \Big)_{\omega = 0}.
\end{eqnarray}
Therefore, the known data will be the radius $R$ and mass $M$ (see \cite{eddington1926}).

In the same way, in the $f(\mathcal{R})$ theory, the radius and the mass of the star will be the same (they are data), so we defined again
\begin{eqnarray} \label{eq:19}
R = (r)_{\omega=0},~~~ GM = \Big(- r^2 \frac{d \phi}{d r} \Big)_{\omega =0}.
\end{eqnarray}
In dimensionless variables we write the equation \eqref{eq:18} with corrections, obtaining
\begin{eqnarray} \label{eq:20}
R'_{f(\mathcal{R})}= (z)_{\omega= 0} = z_{(n)}, 
\end{eqnarray}
\begin{eqnarray} \label{eq:21}
M'_{f(\mathcal{R})} = \Bigg[ \Big(-z^2 \frac{d \omega}{d z} \Big) + \frac{1}{3 m^2 \xi^2} \Big(-z^2 \frac{d \omega^n}{d z} \Big)\Bigg]_{\omega = 0}.
\end{eqnarray}
Thus $R'_{f(\mathcal{R})}$ and $M'_{f(\mathcal{R})}$ are corrections to $R'$ and $M'$ due to modified gravity, and $R$ and $M$ are the known data.

Therefore, in the $f(\mathcal{R})$-gravity model we can write
\begin{eqnarray} \label{eq:22}
\frac{R}{R'_{f(\mathcal{R})}} =  \sqrt{\frac{[(n+1)k]^n}{4 \pi G } \phi_c^{(1-n)}}.
\end{eqnarray}
Note that in this case we have additional terms of correction due to the $f(\mathcal{R})$-gravity. It is also important to note that the values of $z_{(n)}$ for the Newtonian model and $f(\mathcal{R})$-gravity model described by \eqref{eq:18} and \eqref{eq:20} differ from each other since they were derived from different solutions of Lane-Emden equations (standard equation and modified equation for the $f(\mathcal{R})$ theory).

The mass $M(z)$ interior to $z$ is given by \cite{chandrasekhar1957}:
\begin{eqnarray} \label{eq:23}
M(z) = \int^{\xi z}_{0} 4 \pi \rho r^2 dr = 4 \pi \xi^3 \rho_c \int^{z}_{0} z^2 \omega^n dz.
\end{eqnarray}
Using the Lane-Emden equation \eqref{eq:15}, we integrate over the entire star, obtaining
\begin{eqnarray} \label{eq:24}
M= 4 \pi \xi^3 \rho_c \Bigg[\Big(-z^2 \frac{d \omega}{ d z} \Big) + \frac{1}{3 m^2 \xi^2}\Big(-z^2 \frac{d \omega^n}{d z} \Big)\Bigg]_{z_{(n)}}.
\end{eqnarray}
Substituting the expression \eqref{eq:14} that defines $\xi$ and the expression \eqref{eq:4}, we have
\begin{eqnarray} \label{eq:25}
M = 4 \pi \Bigg\{\frac{[(n+1)k]^n}{4 \pi G} \phi_{c}^{(1-n)} \Bigg\}^{\frac{3}{2}} \frac{\phi_{c}^{n}}{[(n+1)k]^n} \times \\ \nonumber \Bigg[ \Big(-z^2 \frac{d \omega}{d z} \Big) + \frac{1}{3 m^2 \xi^2} \Big(-z^2 \frac{d \omega^n}{d z} \Big)\Bigg]_{z_{(n)}}.
\end{eqnarray}
Here we define a parameter $\alpha$ as
\begin{eqnarray} \label{eq:26}
\alpha= m \xi = m \sqrt{\frac{[(n+1)k]^n}{4 \pi G} \phi_{c}^{(1-n)}}.
\end{eqnarray}
Actually $\alpha$ is the free parameter of this model related to correction $f_2$ from modified gravity theory. From this definition, we rewrite equation \eqref{eq:25} as 
\begin{eqnarray} \label{eq:27}
GM = \Bigg\{\frac{[(n+1)k]^n}{4 \pi G}\Bigg\}^{\frac{1}{2}} \phi_{c}^{\frac{(3-n)}{2}} \times \\ \nonumber \Bigg[\Big(-z^2 \frac{d \omega}{d z} \Big) + \frac{1}{3 \alpha^2} \Big(-z^2 \frac{d \omega^n}{d z} \Big)\Bigg]_{z_{(n)}}.
\end{eqnarray}
According to \eqref{eq:21}, we can rewrite \eqref{eq:27} as
\begin{eqnarray} \label{eq:28}
\frac{GM}{M'_{f(\mathcal{R})}} = \Bigg\{\frac{[(n+1)k]^n}{4 \pi G}\Bigg\}^{\frac{1}{2}} \phi_{c}^{\frac{(3-n)}{2}} .
\end{eqnarray}
Here $M$ is the mass data of the star and $M'_{f(\mathcal{R})}$ is the correction calculated from the Lane-Emden solution.

The central condensation is defined as the ratio between the central density of the configuration and its mean density. Thus through the mean density of a star of radius $R = \xi z_{(n)}$ and the expression \eqref{eq:24} for mass, we obtain
\begin{eqnarray} \label{eq:29}
\frac{\rho_c}{\bar{\rho}} = \Bigg(-\frac{3}{z} \frac{d \omega}{d z} - \frac{1}{\alpha^2} \frac{1}{z} \frac{d \omega^n}{d z} \Bigg)_{z_{(n)}}^{-1}.
\end{eqnarray}
In order to write $\phi_c$ as a function of mass and radius, we multiply \eqref{eq:22} and \eqref{eq:28}, obtaining
\begin{eqnarray} \label{eq:30}
\frac{GM}{M'_{f(\mathcal{R})}} \frac{R'_{f(\mathcal{R})}}{R} = \phi_c.
\end{eqnarray}
Thus, by replacing \eqref{eq:29} and \eqref{eq:30} into \eqref{eq:5} in the central region (where $z=0$) and through the definition of $\omega$ by \eqref{eq:13} and the polytropic equation of state \eqref{eq:2}, we find an expression for the pressure
\begin{eqnarray} \label{eq:32}
p &=& \frac{\bar{\rho}}{(n+1)} \frac{GM}{M'_{f(\mathcal{R})}} \frac{R'_{f(\mathcal{R})}}{R} \\ &\times& \nonumber  \Bigg(-\frac{3}{z} \frac{d \omega}{d z} - \frac{1}{\alpha^2} \frac{1}{z} \frac{d \omega^n}{d z} \Bigg)_{z_{(n)}}^{-1}~\omega^{(n+1)}.
\end{eqnarray}
The central pressure $p_c$ is defined for $\omega=1$.

To determine the central temperature of the configurations, we consider the ideal gas law and the equation \eqref{eq:13}, obtaining
\begin{eqnarray} \label{eq:35}
T = \frac{p_c \mu m_{\mu}}{\rho_c k_B} \omega, 
\end{eqnarray}
where $k_B$ is the Boltzmann constant, $\mu$ is the atomic mass and $m_{\mu}$ is the atomic mass unit. 
Using \eqref{eq:5} and \eqref{eq:30}, we write \eqref{eq:35} as
\begin{eqnarray} \label{eq:37}
T = \frac{G \mu m_{\mu}}{(n+1)k_B} \frac{M}{M'_{f(\mathcal{R})}} \frac{R'_{f(\mathcal{R})}}{R} ~\omega,
\end{eqnarray}
from which we define the central temperature $T_c$ when $\omega=1$. It is observed here that for $\alpha \rightarrow \infty$, i.e $m \rightarrow \infty$ and $f_2 \rightarrow 0$ (keeping $\xi$ constant), we recover the stellar structure equations for Newtonian model (see \cite{chandrasekhar1957, eddington1926}).

\section{\label{sec:level4}The polytropic solutions}

In this section, we obtain the polytropic solutions for the stellar structure model according to Newtonian theory and $f(\mathcal{R})$-gravity theory for the following classes of stars: Neutron stars, Brown Dwarfs, White Dwarfs, Red Giants and Sun. We present the results for pressure, temperature and density inside the star according to the value of $n$ and the value of $\alpha$ which is related to $f_2$.  

The curves that describe the behavior of the pressure inside the stars were obtained through expression \eqref{eq:32} for the $f(\mathcal{R})$ theory and the same expressions with $f_2=0$ to the Newtonian model (see \cite{chandrasekhar1957, eddington1926}). We solve numerically the Lane-Emden equation for the respective cases in order to obtain $\omega$ as a function of $z$, consequently $p$ as a function of $z$, and finally, as a function of the distance $r$ from center to surface (where $r=R$) of the star. It is worth to note that in order to compare the curves, for all cases, we have normalized the radius.

To represent a Neutron star in the solution $n=1$, we choose a star of mass $M = 2.01 M_\odot$ and radius $R= 1.87 \times 10^{-5} R_\odot$ as \emph{PSR J0348+0432} \cite{antoniades2013}. In the same way, we obtain the pressure values for a star described by $n=1.5$ of mass $M = 0.053 M_\odot$ and radius $R = 0.1R_\odot$, exemplified by Brown Dwarf \emph{Teide 1} \cite{rebolo1996}. Similarly, for $n=1.5$ we calculate the pressure for a Red Giant star with mass $M= 1.5 M_\odot$ \cite{ohnaka2013, tsuji2008} and radius $R= 44.2 R_\odot$ \cite{richichi2005}, represented by \emph{Aldebaran}. For $n=3$, we obtained the pressure in a White Dwarf of mass $M= 1.5 M_\odot$ \cite{teerikorpi2009} and radius $R = 0.008 R_\odot$ \cite{holberg1998}, exemplified by \emph{Sirius B}, and also to the Sun. Below we have the curves with the pressure values at the center and the surface of the stars for some values of $\alpha$. The chosen values for $\alpha$ in this work were made after many computational tests. We verified that for values greater than $5$, the Newtonian theory was always recovered for all cases studied. Thus we set upper limit at $\alpha =5$ and we investigated what would be the values of $\alpha$ which provide results (for pressure, temperature and density) closest to the observational data for the chosen stars. According to these tests for the chosen stars, in general, the values which provide results closest to the observational data were $\alpha = 0.5,\ 1,\ 5$. 
It is also important explain that it is not possible calculate the surface pressure of the star exactly where $\omega = 0$. In this case, according to equation \eqref{eq:32}, when $\omega \rightarrow 0$, we have $p \rightarrow 0$. So in order to solve this problem, we make an approximation. We choose a point near the surface (point where $\omega$ is close to zero, but not zero) to calculate the surface pressure. This does not cause problem, as it was verified in the computational tests, the values obtained for the pressure in this region near the surface does not vary much (small variations are only in decimal digits). Therefore, we consider this approximation to obtain the pressure on surface. The same applies to surface temperature and density.

\begin{figure*}[ht]
\begin{center}
\includegraphics[width=1.8\columnwidth]{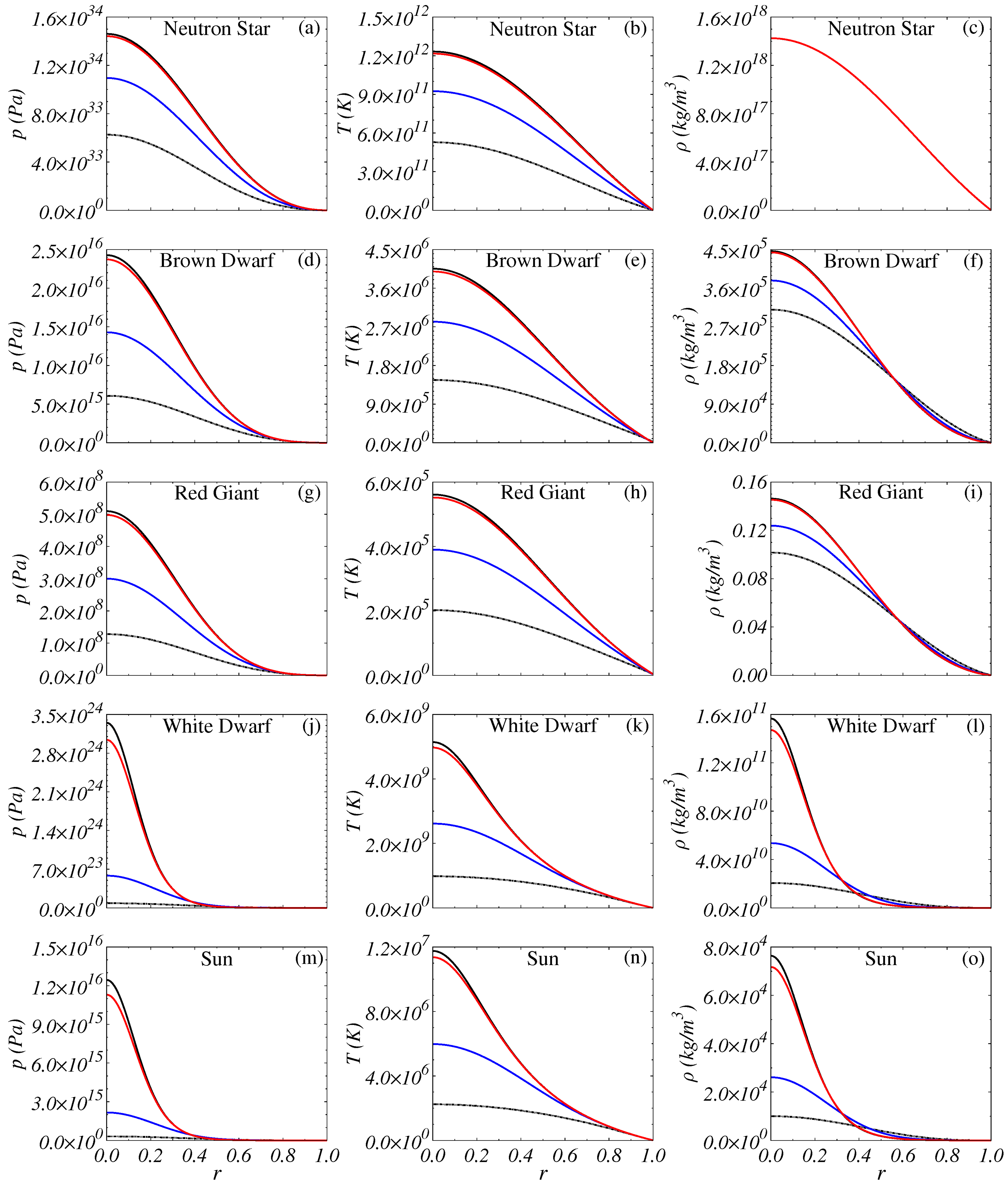}
\caption{\label{Fig.1}Behavior of pressure, temperature and density within some classes of stars (Neutron star, Brown Dwarf, Red Giant, White Dwarf and Sun) describe by Newtonian and $f(\mathcal{R})$-gravity stellar structure models. The $f(\mathcal{R})$-gravity model adopted here considers $f(\mathcal{R})= \mathcal{R} + f_2 \mathcal{R}^2$. The curves in black represent the Newtonian model results, the dot-dashed curves represent the $f(\mathcal{R})$-gravity model results for $\alpha=0.5$, the blue curves for $\alpha=1$ and the red curves for $\alpha=5$.}
\end{center}
\end{figure*}

According to \cite{zhao2015}, the estimated core pressure in a Neutron star is $5.01 \times 10^{34} Pa$, therefore as can be seen through the Fig.\ref{Fig.1}(a) the central pressure results that are closest to this estimate value are those corresponding to the Newtonian model and $f(\mathcal{R})$-gravity model with $\alpha=1$ and $\alpha=5$. In the case of Brown Dwarf, we verify through the Fig.\ref{Fig.1}(d) that central pressure values for $\alpha=1$, $\alpha=5$ and Newtonian model have the same order of magnitude of the expected value to this kind of star, about $10^{16} Pa$ according to \cite{auddy2016}. As can be seen in Fig.\ref{Fig.1}(g), all values obtained for the core pressure in the reported Red Giant have the same order of magnitute expected for a star of that category, approximately $10^8 Pa$. For a White Dwarf, analyzing the Fig.\ref{Fig.1}(j) we conclude that among the values of core pressure found those that best fit the estimate value (approximately $4.95 \times 10^{24} Pa$ in accordance with \cite{teerikorpi2009} are obtained for the Newtonian model and $f(\mathcal{R})$-gravity with $\alpha=5$. Through Fig.\ref{Fig.1}(m) we observe that for the Newtonian model and $\alpha=5$ we find the best results for the pressure in the Sun's core, since according to \cite{williams2016} the estimated value is $2.477 \times 10^{16} Pa$. 

In order to analyze the behavior of the temperature in the same stars, we use the equation \eqref{eq:37} for the $f(\mathcal{R})$-gravity model and the version of these equations for the Newtonian theory \cite{chandrasekhar1957}. It should be remembered that each type of star has an estimated value of atomic mass $\mu$ associated according to its composition (see table \ref{table}).  Also, in order to obtain the temperature as function of $R$ (normalized), we have solved the Lane-Emden equation with the temperature equations mentioned above. Likewise, the surface temperature was calculated considering a $\omega$ value close to it. According to \cite{hong2016}, the expected value for the central temperature of a Neutron star is $1.50 \times 10^{11} K$, therefore observing Fig.\ref{Fig.1}(b) we conclude that the results closest to this value are obtained for $\alpha =0.5$ and $\alpha=1$. Unfortunately, satisfactory data for surface temperature of this star were not found in the literature. As can be seen in Fig.\ref{Fig.1}(e), the results that best fit the expected central and surface temperature values for this Brown Dwarf are those values correspond to $\alpha=1$, since the surface temperature is $2700 K$ and the core temperature is $2.7 \times 10^6 K$, according to \cite{auddy2016}. 

As can be seen in the Fig. \ref{Fig.1}(h), the central temperatures in this Red Giant have the same order of magnitude for all $\alpha$ values considered. However, it is not known an exact value for the temperature in the center of a star like that, since the data found in the literature diverge. Therefore it was not possible to confront in a secure way the values obtained for the models portrayed in this work. The surface temperature in this Red Giant, according to its spectral classification, is about $3.8 \times 10^3 K$ \cite{gray2006}. Thus we find that the best values are those related to $\alpha = 5$ and Newtonian model. In the case of White Dwarf, we verify from Fig.\ref{Fig.1}(k) that all values for the central temperature are one or two order of magnitude higher than expected for this star. In compliance with \cite{teerikorpi2009} the estimated surface temperature is $2.52 \times 10^4 K$ and estimated central temperature is $2.20 \times 10^7 K$. For the surface temperature, the value relative to $\alpha=1$ performs the best result. According to \cite{williams2016} the estimated surface temperature of the Sun is $5780 K$ and its central temperature is $1.571 \times 10^7 K$. Thus, as can be seen in Fig.\ref{Fig.1}(n), among the values obtained for the temperature in the Sun's core, those best fitted to the data are related to $\alpha =5$ and Newtonian model. For the surface temperature, all values are compatible with the data. 
\begin{table}
\begin{center}
\caption[]{Estimates of atomic mass and composition (mass fraction) for each type of star. The mass fraction of Hydrogen is represented by $X$, the mass fraction of Helium by $Y$ and the heavy metals by $Z$ \cite{wilking1999, cercignani2002, asplund2009}.}\label{table}
\begin{tabular}{clcl}
  \hline\noalign{\smallskip}
Star & Composition (Mass Fraction) &  $\mathnormal{\mu}$  \\
  \hline\noalign{\smallskip}
Neutron Star & Neutrons & 1.00 \\
Brown Dwarf & X =0.70, Y=0.28, Z=0.02 & 0.62   \\ 
White Dwarf & X=0, Y=0, Z=1 & 2.00 \\
Red Giant & X=0, Y=0.98, Z=0.02 & 1.34   \\
Sun & X=0.73, Y=0.25, Z=0.02 & 0.60  \\
  \noalign{\smallskip}\hline
\end{tabular}
\end{center}
\end{table}

To determine the behavior of the matter density in the stellar interior, we solve the equation \eqref{eq:29} considering the mean density calculated through the mass $M$ and the radius $R$ of the star. Undoubtedly it is expected in all situations that the density will decrease as it approach the surface as observed with temperature and pressure. The density at surface can obtained considering a point near the interface of the star with the outside where $\omega$ vanishes, according to the boundary conditions adopted. For Neutron star (see Fig. \ref{Fig.1}(c)), we have the case where all the curves coincide, so the values obtained for the central and superficial density do not present considerable differences. Therefore, in this situation, all values of central density are close to the reported value for a Neutron star with the aspects described above, according to \cite{zhao2015} the expected value of the core density is $1.50 \times 10^{18} kg/m^3$. In the case of Brown Dwarf, the central density is estimated between $10^3 kg/m^3$ and $10^6 kg/m^3$. Therefore, through Fig.\ref{Fig.1}(f), we conclude that the core density resulting from the models discussed in this work presents values within order of magnitude expected for a Brown Dwarf with the characteristics mentioned previously \cite{burrows1993, rebolo1996}. For Red Giants the core densities, shown in Fig. \ref{Fig.1}(i), have the same order of magnitude. Again, the data found in the literature for this star are not congruent, which makes it difficult to compare with the results obtained. For the central density of White Dwarf, according to Fig. \ref{Fig.1}(l), the best results are those found for $\alpha =0.5$ and $\alpha=1$, since the estimated value is $3.30 \times 10^{10} kg/m^3$ in compliance with \cite{teerikorpi2009}. Observing the Fig.\ref{Fig.1}(o), we verify that the central densities obtained for the Sun presents values with an order of magnitude lower than that assured by the data: $1.622 \times 10^5 kg/m^3$ according to \cite{williams2016}, including the Newtonian polytropic model, considered as a reasonable model for the description of Sun. 

Through the Fig.\ref{Fig.1}, it has been found that in all cases, curves show a more rapid decrease as $\alpha$ decreases. Therefore, the more the model $f(\mathcal{R})$ departs from the Newtonian theory, the lower predicted pressure, temperature and density in the center of the star.  

\section{\label{sec:level5}Conclusions}

In this work we analyzed the stellar structure from the point of view of $f(\mathcal{R})$-gravity. The reason that led us to adopt such an approach is the fact that higher-order curvature corrections can emerge in intense gravitational field regimes, as occurs within the stars. In this scheme it is tangible to assume that the emergence of these corrections may generate effects on pressure, temperature and density, for example. Thus, in the stellar structure model equations, new terms related to the curvature corrections lead to different behaviors of these magnitudes. It is worth mentioning that since these quadratic terms arise in strong field regimes, in the Solar System scale we have the weak field scheme where only the linear terms of Ricci scalar $\mathcal{R}$ are relevant.

In order to perform the analysis of the stellar structure under the focus of this extended theory of gravity, we started with an action that represents the $f(\mathcal{R})$-gravity models, adopting as function $f(\mathcal{R})= \mathcal{R} + f_2 \mathcal{R}^2$, the hydrostatic equilibrium equation and the polytropic equation of state. These equations are employed in the non-relativistic approximation, making the model a simple description of the complex behavior of these stellar object classes. The goal is to illustrate the effects of the $f(\mathcal{R})$-gravity theory adopted in this work on an easily understandable model. According to this assumption we solved numerically the equations responsable for the stellar structure, evidencing the role that the corrections in the curvature play in these equations. In this way, the expressions for pressure, density and temperature depend strictly on the values of these corrections. Interpretating these additional terms as corrections of General Relativity, we control deviations from the model with respect to Einstein's theory. It is worth mentioning that in the solutions from the $f(\mathcal{R})$-gravity model, we noticed the presence of an extra degree of freedom $\alpha$ related to the correction $f_2$.

Through the Fig. \ref{Fig.1}(a) - \ref{Fig.1}(o), temperature, pressure and density of different classes of stars, described by Newtonian theory and $f(\mathcal{R})$-gravity model, were compared numerically. These quantities were plotted against the normalized radius of the star. Observing these graphics, we found that by increasing the correction term $f_2$, consequently decreasing $\alpha$ (according to equation \eqref{eq:26}, keeping $\xi$ constant, $\alpha$ and $f_2$ are inversely proportional), these physical quantities within the star exhibit a faster decrease and lower core values. In this way, the more extended theory moves away from the Newtonian theory (more $\alpha$ decreasing), the lower the pressure and the temperature inside the star. With respect to density, for $n=1$ the curves coincide for all values of $\alpha$ chosen, including for the Newtonian case. For the other values of $n$, we observed the same behavior that was described for the pressure and the temperature, the lower $\alpha$, the lower the central density of the star and more faster the curve decreases. All this leads us to belive that the magnitude of gravitational corrections changes the stellar structure. Therefore, by increasing the value of parameter $f_2$, the original pressure, density and temperature of General Relativity are affected by a term that modifies the mass and radius of the star in some way. It is important to emphasize that we did not use exotic matter, only the usual matter (baryonic). We adopted the \emph{Jordan frame}, so that the results of the gravitational sector were corrected while the matter sector was not affected. Thus, the geodesic structure was not changed and the standard polytropic state equation could be assumed.

By analyzing the graphics it can be seen that, in general, the best description of Neutron stars, Brown Dwarfs and White Dwarfs was achieved by the parameter $\alpha=1$. This fact corroborates the expected behavior of the physical quantities in stars with intense gravitational fields. It is worth mentioning that this value of $\alpha$ represents a case in which curvature is more accentuated and consequently, the gravitational field is more intense. Therefore, in this work we show that, in these cases, for a better description is needed a model of stellar structure according to the $f(\mathcal{R})$-gravity theory, since the contribution of the corrections in the quantities is more expressive. Another issue to be discussed is based on the analysis of the results obtained for Neutron stars. For these stars, some results found did not show significant differences between the values obtained by the $f(\mathcal{R})$ and Newtonian models, although the model with $\alpha=1$ approached more than the observed values for this star. Differences as expressive as those observed in the White Dwarfs were expected, since the Neutron stars also have intense gravitational fields. We justify this fact by noting that the equation of state used here (polytropic equation) is not the most adequate, since it does not consider the quantum effects present in this type of star. The Red Giants and the Sun (stars with weak fields) were better represented by Newtonian model and $\alpha=5$ (in all cases reported in this work, the curves corresponding to this parameter value are almost coincident to Newtonian's curve). This reinforces a fact observed in the graphics: for values $\alpha=5$, we have already been able to recover the Newtonian description for the stars. In this way, we can conclude through this work that in the case where the associated gravitational field is weak, the best description is obtained through a Newtonian model, not being necessary the approach through a $f(\mathcal{R})$ theory.

As presented, this work was intended to indicate the possibility of describing some classes of stars in polytropic models under a different assumption about gravity. The study of these systems in this approach may be important to testing $f(\mathcal{R})$-gravity theories, since strong gravitational field regimes are located in stars. Despite the simplicity of the model, the results were satisfactory. The estimated values for pressue, density and temperature are within those determined by observations. The results of this work can be extended to stars with magnetic and rotating fields, for example, and for different equation of state.

We also emphasize that the results obtained in this work could not be compared with others results from other articles, incluinding those cited throughout this manuscript, except in the case of Newtonian theory, whose comparasion was performed and presented here. Until the moment, all papers found in the literature have different motivations and objectives from those exposed by us.

To sum up we have determined the fields of density, pressure and temperature for stars by using a $f(\mathcal{R})$-gravity model and compared with the results that came out from the Newtonian theory. The stars analyzed were of two types: with  strong fields such as White Dwarfs, Neutron stars and Brown Dwarfs and with weak fields such as Red Giants and Sun. The $f(\mathcal{R})$-gravity model has been proved to be necessary for the description of the stars with strong fields, and as was expected the Newtonian theory provides a good description for stars with weak fields. 

\begin{acknowledgments}
The authors acknowledge the financial support of CNPq (Brazil).
\end{acknowledgments}

\bibliography{bibtex}

\end{document}